\documentclass[pra,aps,twocolumn,showpacs,10pt]{revtex4-1}
\usepackage{amsmath,amssymb,graphicx}
\usepackage{amsmath,amssymb,amsthm,dsfont,bm,color,ulem}

\begin{document}

\title{Stabilization of vortex beams in Kerr media by nonlinear absorption}
\author{Miguel A. Porras$^1$}
\author{M\'arcio Carvalho$^1$}
\author{Herv\'e Leblond$^2$}
\author{Boris A. Malomed$^{3,4}$}

\begin{abstract}
We elaborate a new solution for the problem of stable propagation of
transversely localized vortex beams in homogeneous optical media with
self-focusing Kerr nonlinearity. Stationary nonlinear Bessel-vortex states
are stabilized against azimuthal breakup and collapse by multiphoton
absorption, while the respective power loss is offset by the radial influx
of the power from an intrinsic reservoir. A linear stability analysis and
direct numerical simulations reveal a region of stability of these vortices.
Beams with multiple vorticities have their stability regions too. These
beams can then form robust tubular filaments in transparent dielectrics as
common as air, water and optical glasses at sufficiently high intensities.
We also show that the tubular, rotating and speckle-like filamentation
regimes, previously observed in experiments with axicon-generated Bessel
beams, can be explained as manifestations of the stability or instability of
a specific nonlinear Bessel-vortex state, which is fully identified.
\end{abstract}


\affiliation{$^1$Grupo de Sistemas Complejos, Universidad Polit\'ecnica de Madrid, Rios Rosas 21, 28003 Madrid, Spain}
\affiliation{$^2$Laboratoire de Photonique d'Angers, EA 4464, Universit\'e d'Angers, 2 Bd Lavoisier, 49000 Angers, France}
\affiliation{$^3$Department of Physical Electronics, School of Electrical Engineering, Faculty of Engineering, Tel Aviv University, Tel Aviv 69978, Israel}
\affiliation{$^4$Laboratory of Nonlinear-Optical Informatics, ITMO University, St. Petersburg 197101, Russia}

\maketitle

\section{Introduction}

Self-trapping of two-dimensional (2D) localized modes (spatial solitons) in nonlinear diffractive media has been one of central topics in nonlinear
optics, with many ramifications to other areas \cite{general-reviews,KA,RMP}. Belonging to this species of self-trapped modes is the first example of
solitons introduced theoretically in nonlinear optics, \textit{viz}., the \textit{Townes soliton} \cite{Townes} supported by the cubic (Kerr)
self-focusing nonlinearity. Later, counterparts of the Townes soliton with embedded vorticity, i.e., tubular vortical beams, were theoretically
introduced \cite{Minsk}. Vorticity-carrying hollow beams may find applications as conduits guiding weak optical signals in all-optical
data-processing schemes \cite{KA}, as well as tweezers trapping small particles and setting them in rotation \cite{tweezers}.

Unlike 1D solitons, which are usually stable objects \cite{KA,Michel}, stability is a major issue for their 2D counterparts. The common cubic
self-focusing nonlinearity, which easily supports formal 2D solitons solutions, gives rise to the \textit{critical collapse} \cite{collapse},
which destabilizes the soliton families. For this reason, the Townes soliton has never been observed in the experiment. Multidimensional solitons with embedded vorticity, alias vortex rings, are vulnerable to a still stronger instability initiated by azimuthal perturbations, which splits the ring into fragments \cite{general-reviews}.

Thus, the stabilization of multidimensional solitons is an issue of great significance to theoretical studies, and creation of the solitons in the
respective physical settings is a challenge to the experiment. The stabilization of fundamental and vortex solitons by harmonic-oscillator
trapping potentials, which do not break the underlying spatial isotropy, has been theoretically elaborated in detail \cite{2D,Dum}. Vortex stabilization was shown to be possible with specific nonlinear terms, such as cubic and quintic ones with opposite signs \cite{MICHINEL1,MICHINEL2}, or nonlocal nonlinearities \cite{KRUGLOV,YAKIMENKO}. Their practical implementation requires a careful search for suitable materials, such as liquid CS$_2$ for the cubic-quintic nonlinearity \cite{ARAUJO}, or lead-doped glass for the thermal nonlocal nonlinearity \cite{ROTSCHILD}. On the other hand, a recently proposed method applying to binary systems with spin-orbit coupling between the components, provides the stabilization of solitons, combining fundamental and vortical terms in their two components, in both the 2D \cite{we} and 3D \cite{HP} settings. A related topic is the stability of fundamental and vortex solitons in dissipative media. Most often, complex Ginzburg-Landau equations (CGLEs) with the cubic-quintic nonlinearity are used to predict stable dissipative solitons and spiral vortices \cite{spiral,CQ}.

A different scenario, closely connected to the phenomenon of filamentation, was considered in Ref. \cite{PORRAS1}, which addressed the stationary
propagation of nonlinear Bessel beams in media combining self-focusing Kerr nonlinearity and nonlinear dissipative terms accounting for multiphoton
absorption (MPA). The occurrence of MPA of different orders in usual dielectric media, such as silica, water, or air, is a well-established fact.
MPA is a collapse-arresting mechanism, which is an essential ingredient in the filamentation induced by powerful ultrashort pulses \cite{Shimshon,COUAIRON,Bessel-beams,PORRAS1,POLESANA,POLESANA2}. Being weakly localized, these nonlinear beams carry an unlimited amount of power, playing the role of the intrinsic reservoir which supports an inward-directed radial power flux balancing the nonlinear absorption around the beam's center. A drastic difference from the CGLE-based models is that the propagation of these Bessel-like beams in media with nonlinear losses may persist without any gain. Of course, in a real experiment the power stored in any beam is finite, but the propagation distance needed for its depletion may easily exceed the length of the experimental sample, thus making the model quite realistic. Nonlinear Bessel beams play a crucial role in various experiments \cite{POLESANA,POLESANA2}, including filamentation with ultrashort pulsed Bessel beams \cite{COUAIRON}, where their nonlinear counterparts act as stable or chaotic attractors \cite{PORRAS2}. A qualitatively similar situation was demonstrated in an experiment with influx of atoms into a
dissipative sink embedded into a quasi-infinite reservoir \cite{Ott}.

More recently, vorticity-carrying nonlinear Bessel modes, alias Bessel vortex beams (BVBs), with a hollow transverse structure, were introduced
theoretically \cite{PORRAS3,JUKNA}. Similarly to the fundamental nonlinear Bessel beam \cite{PORRAS1}, and also to nonlinear Airy beams \cite{LOTTI},
stationarity is supported by an inward-directed power flux balancing the losses. Subsequently, BVBs were realized experimentally in glass, and
employed for laser-powered material processing \cite{XIE}. Depending on the input power and geometry, different propagation regimes featuring tubular, rotating and non-rotating speckle-like filaments were observed within the limits of the so-called Bessel distance associated to the depletion of the
reservoir \cite{JUKNA,XIE}.

In this paper we first show that BVBs may be \textit{completely stable} against the azimuthal breakup and collapse alike, in the medium combining
the focusing Kerr nonlinearity and MPA. The stability, which is actually imposed by the MPA effect, is predicted by the analysis of the linearized
equations for small perturbations, and verified in direct numerical simulations. The stabilizing effect of nonlinear absorption is known in
other contexts, such as the stabilization against radial and temporal perturbations of zero-vorticity beams \cite{POLESANA2,PORRAS2} and light
bullets \cite{PORRAS4}, but not against the azimuthal breaking, to which nonlinear vortex beams are vulnerable. We focus on the ubiquitous cubic
self-focusing, which, by itself, cannot support any stable 2D patterns. Similar results can be readily obtained for more general nonlinearities,
since the existence of the BVBs does not critically depend on the precise nonlinearities, such as the order of MPA, or the precise from of the Kerr
nonlinearity (cubic, cubic-quintic\dots) \cite{PORRAS1,PORRAS3,JUKNA}. Thus, we predict that stable BVBs can be created in almost any transparent
dielectric at high enough intensities.

In the light of the linear-stability analysis, we perform diagnostic numerical simulations of axicon-generated vortex Bessel beams propagating in
nonlinear media, that allows us to propose a common explanation of the three observed dynamical regimes \cite{XIE}. In all the three cases, there exist a BVBs that determines the dynamics in the Bessel zone after the axicon. This BVB is specified unambiguously as one preserving three characteristics of the linear Bessel beam that would be formed about the middle of the Bessel zone in linear propagation, namely, the cone angle, the topological charge, and the inward radial power flux. The dynamics observed in the Bessel zone corresponds to that expected from the development of instability, if any, of that BVB. If this BVB is stable (according to the linear-stability analysis), or its instability is weak to develop over the Bessel distance, the tubular regime is observed in the Bessel zone; otherwise, the azimuthal symmetry breaking in the development of instability,
leading either to a rotatory or to a speckle-like filament regime, is observed in the Bessel zone.

\section{Nonlinear Bessel vortex beams in Kerr media with multiphoton
absorption}

We consider the transmission of a paraxial light beam whose electric field is $E=\mathrm{Re}\left\{ A\exp [i(kz-\omega t)]\right\} $, with complex
envelope $A$, carrier frequency $\omega $, and propagation constant $k=n\omega /c$ ($n$ is the linear refractive index and $c$ the speed of light
in vacuum), in the Kerr medium with MPA. The evolution of the envelope is governed by the nonlinear Schr\"{o}dinger equation
\begin{equation}
\partial _{z}A=\frac{i}{2k}\Delta _{r}A+i\frac{kn_{2}}{n}|A|^{2}A-\frac{\beta ^{(M)}}{2}|A|^{2M-2}A\,,  \label{NLSE1}
\end{equation}
where $(r,\varphi ,z)$ are cylindrical coordinates, $\Delta _{r}\equiv \partial _{r}^{2}+(1/r)\partial _{r}+(1/r^{2})\partial _{\varphi }^{2}$, $n_{2}>0$ is the nonlinear refractive index, and $\beta ^{(M)}>0$ is the $M$-photon-absorption coefficient. The multiphoton order $M$ is determined by the medium and the light wavelength \cite{COUAIRON2}. In air, for instance, $M$ ranges from $3$ to $8$ in the wavelength range $248$--$800$ nm \cite{SCHWARZ,COUAIRON}.

\begin{figure}[tbp]
\includegraphics[width=4.0cm]{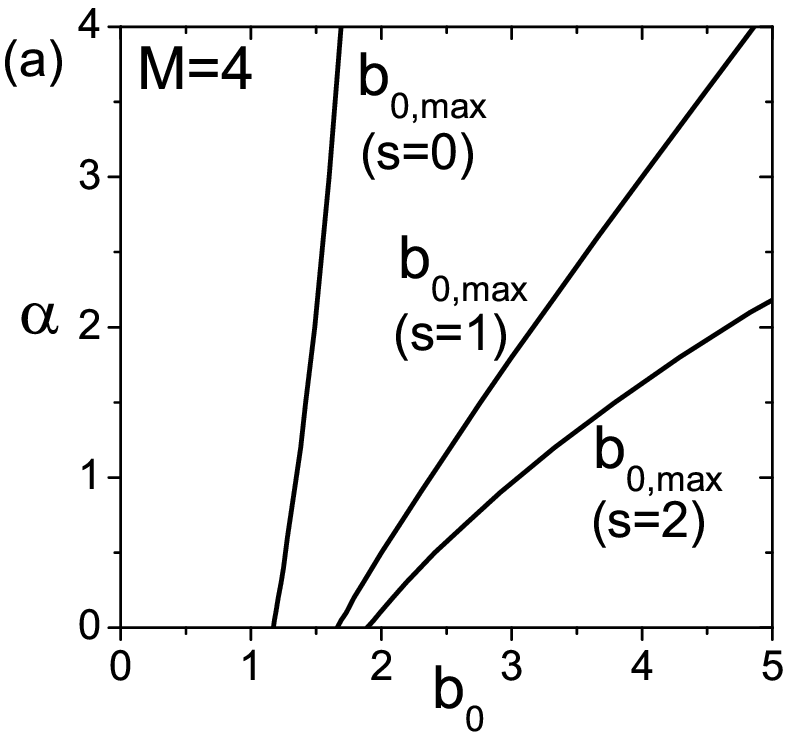}
\includegraphics[width=4.1cm]{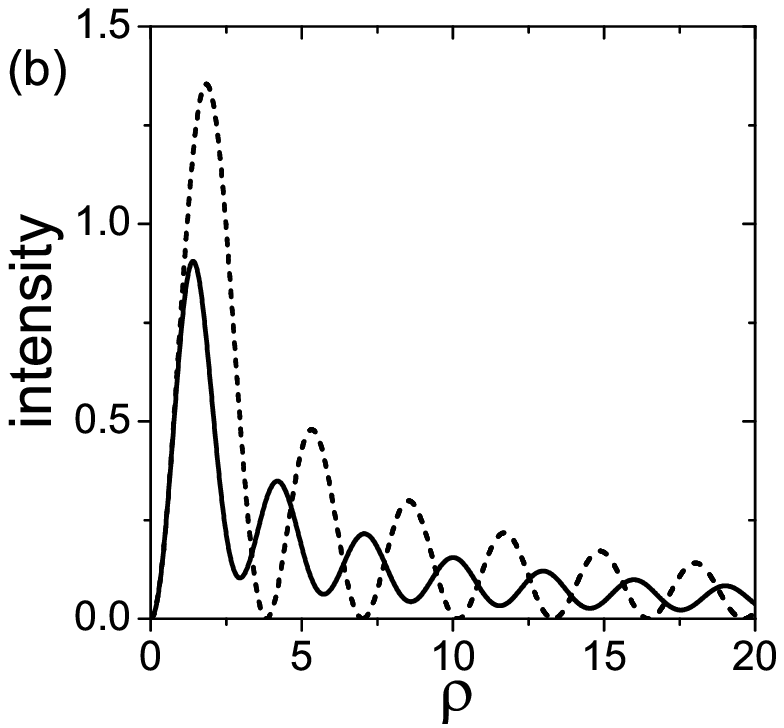}
\caption{(a) For $M=4$, BVBs with $s=0,1$ and $2$ exist at $b_{0}<b_{0,\max}$. The dependence of $b_{0,\max}$ on $\protect\alpha$ defined in Eq. (\protect\ref{alpha}), and relating Kerr nonlinearity, nonlinear absorption and cone angle, is shown in the plot. (b) The dimensionless intensity profile $|\tilde A|^2$ of the nonlinear BVB with $s=1$, $\protect\alpha =1$ and $b_{0}=2$ (solid curve), compared to the intensity profile $b_{0}^{2}J_{s}^{2}(\protect\rho )$ of the linear Bessel vortex with the same radial profile at $r\rightarrow 0$ (dashed curve).}
\label{fig1}
\end{figure}

Nonlinear BVBs are solutions to Eq. (\ref{NLSE1}) of the form \cite{PORRAS3,JUKNA} $A_{s}=a(r)\exp \left[ i\phi (r)+is\varphi +i\delta z\right]
$, where real amplitude $a(r)>0$ vanishes at $r\rightarrow \infty $, $s=0,\pm 1,\pm 2\dots $ is the vorticity, and, on the contrary to the usual
vortex solitons \cite{general-reviews} (which do not exist in this model), the axial wavenumber shift $\delta \equiv -k\theta^{2}/2$ is negative, being associated with the conical structure of the paraxial beam with half-apex angle $\theta$. We define the scaled radial coordinate, propagation
distance and envelope 
\begin{equation}  \label{SCALING}
\rho \equiv k\theta r=\sqrt{2k|\delta |}r, \quad \zeta \equiv |\delta |z,\quad \tilde{A}\equiv \left(\frac{\beta^{(M)}}{2|\delta |}\right)^{\frac{1}{2M-2}}A,
\end{equation}
to rewrite Eq. (\ref{NLSE1}) as
\begin{equation}
\partial _{\zeta }\tilde{A}=i\Delta _{\rho }\tilde{A}+i\alpha |\tilde{A}|^{2}\tilde{A}-|\tilde{A}|^{2M-2}\tilde{A}\,,  \label{NLSE2}
\end{equation}
where
\begin{equation}  \label{alpha}
\alpha \equiv \left( \frac{2|\delta |}{\beta ^{(M)}}\right) ^{1/(M-1)}\frac{k n_{2}}{n|\delta |}>0
\end{equation}
is determined by the medium properties at the light wavelength and the cone angle. For example, in water at $527$ nm, with values $n=1.33$, $n_2=2.7\times 10^{-16}$ cm$^2/$W, $M=4$ and $\beta^{(4)}=2.4\times 10^{-36}$ cm$^5/$W$^3$ \cite{POLESANA2}, and with cone angles $\theta=3^\circ,2^\circ,1^\circ,0.5^\circ$, we obtain the respective values $\alpha\simeq 0.76,1.31,3.30,8.31$. Similar values of $\alpha$ are obtained for other media at other wavelengths, and are therefore considered below. As follows from Eq. (\ref{NLSE2}), the amplitude and phase of the BVB's field, $\tilde{A}_{s}=\tilde{a}(\rho )e^{i\phi (\rho )}e^{is\varphi }e^{-i\zeta }$, satisfy the equations
\begin{gather}
\frac{d^2\tilde a}{d\rho^2}+\frac{1}{\rho}\frac{d\tilde a}{d\rho}-\left(\frac{d\phi}{d\rho}\right)^{2}\tilde{a}+\tilde{a}+\alpha \tilde{a}^{3}-\frac{s^2}{\rho^2}\tilde{a}=0\,,  \label{AP1} \\
\frac{d^2\phi}{d\rho^2}+\frac{1}{\rho}\frac{\phi}{d\rho}+\frac{2}{\tilde a}\frac{d\phi}{d\rho}\frac{d\tilde a}{d\rho} +\tilde{a}^{2M-2}=0\,,
\label{AP2}
\end{gather}
In the absence of nonlinearities, these equations give rise to the linear Bessel beam $\tilde{a}(\rho)e^{i\phi (\rho )}=b_{0}J_{s}(\rho )$ with
arbitrary $b_{0}>0$. In the presence of the nonlinear terms, nonlinear BVBs with the same vortex core, i.e., $\tilde{a}(\rho )e^{i\phi (\rho )}\simeq
b_{0}J_{s}(\rho )$ at $\rho $ small enough, exist up to a maximum value of $b_{0}$, $0<b_{0}<b_{0,\mathrm{max}}$, where $b_{0,\mathrm{max}}$ depends on $\alpha $, as shown in Fig. \ref{fig1}(a). Thus, for the given MPA order, $M$, and topological charge $s$, the BVB is specified by the two parameters, $\alpha $ and $b_{0}$. A typical nonlinear BVB intensity profile is shown in Fig. \ref{fig1}(b). Compared to the linear Bessel vortex with the same $b_{0}$, the rings increasingly compress with the increase of $\alpha $, and feature a reducing contrast as $b_{0}$ approaches $b_{0,\mathrm{max}}$. The parameter $b_{0}$ also controls the peak intensity of the first ring, which is a growing function of $b_{0}$ for fixed $\alpha$. Values of $b_0$ of the order of unity describe typical intensities in experiments where the interplay between self-focusing and nonlinear absorption was observed to play an essential role \cite{POLESANA2}. For example, with $M=4$, $s=1$ and $\alpha=3.30$, values $b_0=1.2$ and $1.6$ pertain, respectively, to BVBs with cone angle $\theta=1^\circ$ and peak intensities $0.77$ TW/cm$^2$ and $1.16$ TW/cm$^2$ in water at $527$ nm.

Asymptotically at large radius $\rho$, a nonlinear BVB behaves as a superposition of two H\"{a}nkel beams of different amplitudes:
\begin{equation}
\tilde{A}_{s}\simeq \frac{1}{2}[b_{\mathrm{out}}H_{s}^{(1)}(\rho )+b_{\mathrm{in}}H_{s}^{(2)}(\rho )]e^{is\varphi }e^{-i\zeta },
\end{equation}
where the amplitudes of the outward and inward H\"ankel beams satisfy $|b_{\mathrm{in}}|^{2}-|b_{\mathrm{out}}|^{2}=2\pi \int_{0}^{\infty }d\rho \rho
\tilde{a}^{2M}(\rho )$ to maintain the equality of the net inward-directed power flux, $|b_{\mathrm{in}}|^{2} - |b_{\mathrm{out}}|^{2}$, and the
power-loss rate in the beam's core. The values of the amplitudes $|b_{\mathrm{in}}|$ and $|b_{\mathrm{out}}|$ for each BVB can easily be extracted
form an analysis of its transversal intensity profile, as explained in detail in \cite{PORRAS3}.

We point out that the existence of these BVBs and their properties do not critically depend on the optical properties of the medium, such as the MPA
order $M$, the coefficient $\beta^{(M)}$, and the nonlinear index $n_2$. These beams are also supported in self-defocusing media, as well as in those
with the cubic-quintic nonlinearity \cite{PORRAS3}. Only details of transversal profile, the range of existence $b_{0,\mathrm{max}}$, and amplitudes of the H\"ankel components differ from case to case.

\section{Stability analysis for single and multiple vortices, and numerical verification}

Following the usual approach to the linear-stability analysis \cite{general-reviews}, we take a perturbed nonlinear BVB as $\tilde{A}=\tilde{A}
_{s}+\epsilon \left[ u_{m}(\rho )e^{i\kappa \zeta +im\varphi }+v_{m}^{\star}(\rho )e^{-i\kappa ^{\star }\zeta -im\varphi }\right] e^{is\varphi -i\zeta }$, where $\epsilon $ is an infinitesimal amplitude of perturbations with eigenmodes $[u_{m}(\rho ),v_{m}(\rho )]$ and integer winding number $m$. The instability takes place if there exists at least one eigenvalue with $\kappa_{\mathrm{I}}\equiv \mathrm{Im}\{\kappa \}<0$. The linearized
equations obtained by substituting this Ansatz in Eq. (\ref{NLSE2}) are
\begin{equation}
\left(
\begin{array}{cc}
H_{+} & f \\
-f^{\star } & -H_{-}^{\star }
\end{array}
\right) \left(
\begin{array}{c}
u_{m} \\
v_{m}%
\end{array}
\right) =\kappa \left(
\begin{array}{c}
u_{m} \\
v_{m}
\end{array}
\right) ,  \label{EIGEN}
\end{equation}
where $f\equiv \lbrack \alpha \tilde{a}^{2}+i(M-1)\tilde{a}^{2M-2}]e^{2i\phi}$, and
\begin{equation}
H_{\pm }\equiv \frac{d^{2}}{d\rho ^{2}}+\frac{1}{\rho }\frac{d}{d\rho }-\frac{(s\pm m)^{2}}{\rho ^{2}}+1+(2\alpha \tilde{a}^{2}+iM\tilde{a}^{2M-2}).
\label{H}
\end{equation}
Solutions for $u_{m}(\rho )$ and $v_{m}(\rho )$ must vanish at $\rho \rightarrow \infty $, being subject to the usual boundary conditions at $\rho \rightarrow 0$: $u_{m}\sim \rho ^{|s+m|}$, $v_{m}\sim \rho ^{|s-m|}$. In the case of instability, the brightest ring, and possibly the secondary
one too, of a weakly perturbed BVB are expected to break into fragments, the number of which is equal to winding number $m$ of the mode with the largest growth rate \cite{general-reviews}.

\begin{figure}[tbp]
\begin{center}
\includegraphics[width=4cm]{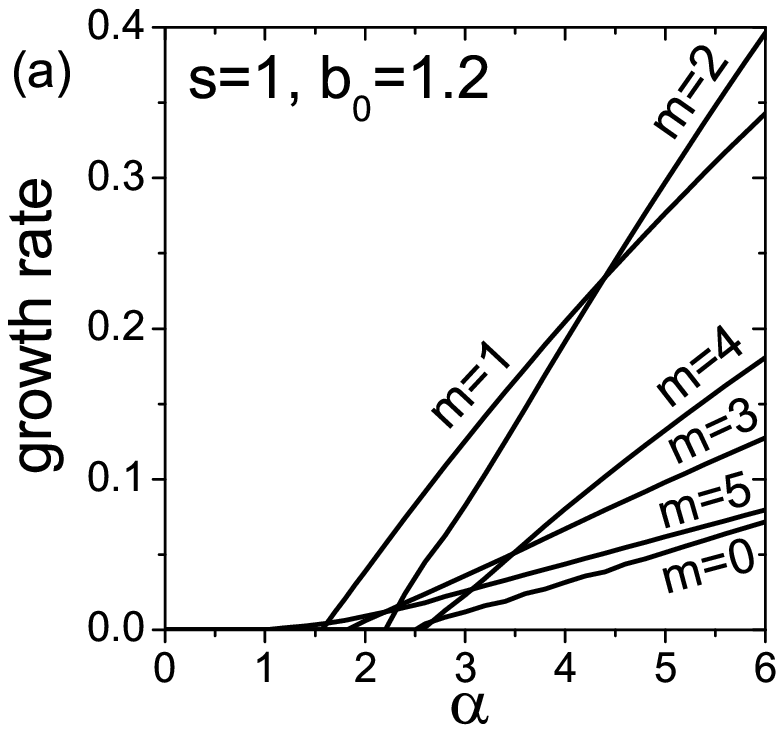} \includegraphics[width=4cm]{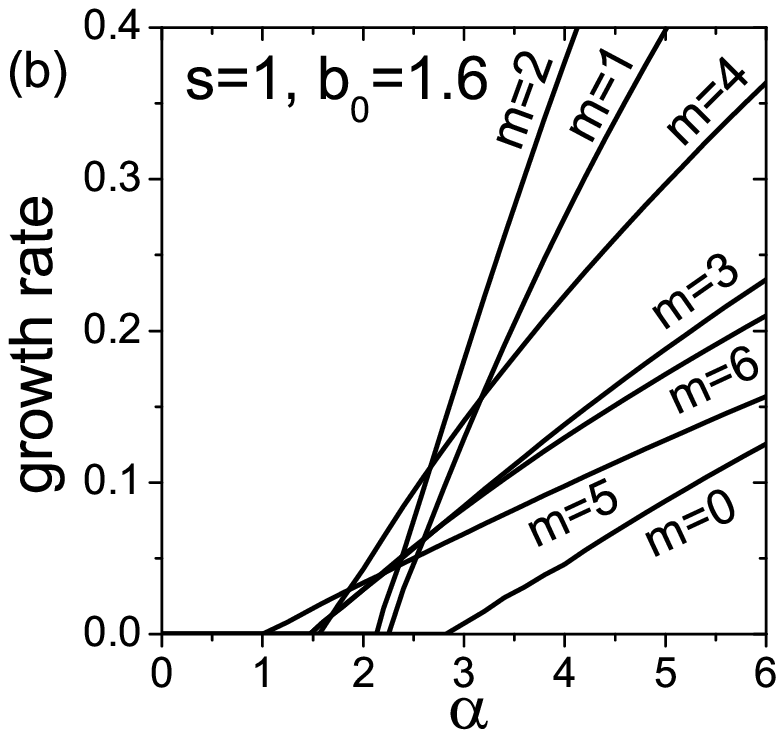}
\includegraphics[width=4cm]{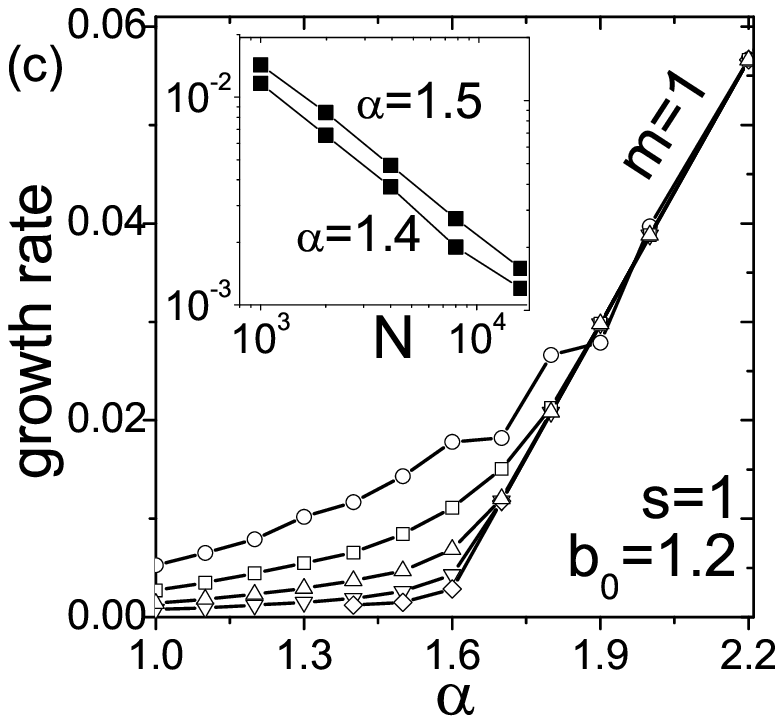} \includegraphics[width=4cm]{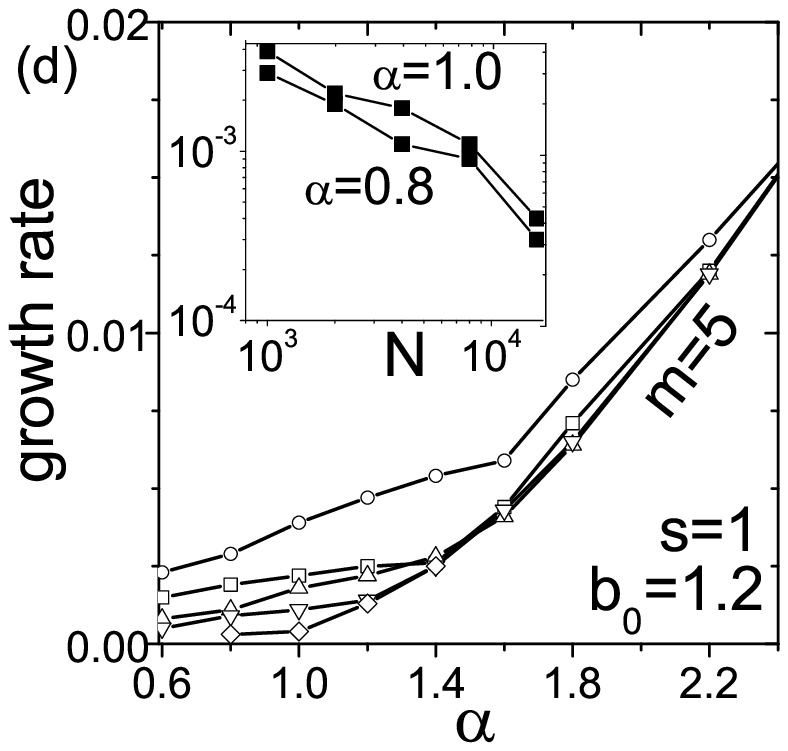}
\end{center}
\caption{(a,b) Dimensionless Growth rates of unstable perturbation modes versus $\protect\alpha$, for values of $s$, $b_{0}$ and $m$ indicated in the panels. (c,d) The largest growth rate computed with stepsize $h=0.2$ and increasing numbers of points in the numerical grid, $N=1000$ (circles), $2000$ (squares), $4000$ (up triangles), $8000$ (down triangles) and $16000$ (rhombuses), for values of $\protect\alpha$ at which the computed growth
rates significantly depend on $N$. For all perturbation winding numbers $m$, the growth rates at large $\protect\alpha$ quickly converge to positive
values as $N$ increases, thus representing a genuine instability of the underlying BVB. Instead, for smaller values of $\protect\alpha$ the growth
rates approach zero as $N$ increases, implying that the BVBs are stable in this case. The insets additionally display the growth rates as functions of $N$ for particular values of $\protect\alpha $, indicating a decay $\sim 1/N$.}
\label{fig2}
\end{figure}

The numerical procedure for solving Eq. (\ref{EIGEN}) is more difficult for BVBs than for familiar vortex solitons in CGLEs \cite{spiral,CQ} due to the weaker localization. Figures \ref{fig2}(a) and (b) show the instability growth rates for the BVBs with $s=1$, two different values of $b_{0}$ and
varying $\alpha$. Figures \ref{fig2}(c) and (d) illustrate how these results are obtained from the eigenvalue spectrum of Eq. (\ref{EIGEN}) with radial derivatives discretized on a radial mesh of $N$ points of stepsize $h$, and zero boundary condition at the truncation radius, $\rho _{\mathrm{trun}}=Nh$, in the limit of $h\rightarrow 0$ and $N\rightarrow \infty $. No substantial difference between the eigenvalues for different small values of $h$ was found at fixed $Nh$ provided that the BVB profile is adequately sampled. In Figs. \ref{fig2}(c) and (d) we then focus on effects of increasing $Nh$ by fixing $h$ and increasing $N$. We note that weakly unstable modes produced by Eq. (\ref{EIGEN}) feature the slow exponential decay $u,v\sim e^{\kappa _{I}\rho /2}$ \cite{POLESANA2}, hence they can be suitably reproduced by the truncated system if $Nh\gg 1/|\kappa_{I}|$. In other words, the smallest reliable growth rate produced by the truncated system is estimated as $\sim 1/Nh$. To improve the accuracy for these small growth rates, we used special routines to obtain eigenvalues of sparse matrices up to $32000\times 32000$ for $N=16000$. The results of this analysis, shown in Figs. \ref{fig2}(a) and (b), allow us to conclude that there exist BVBs that are \textit{truly stable} against small perturbations (e.g., for $M=4$, and $s=1$ these are the BVBs with $\alpha \leq 1.1$ for $b_{0}=1.2$, and $\alpha \leq 1$ for $b_{0}=1.6$).

\begin{figure}[tbp]
\begin{center}
\includegraphics[width=4cm]{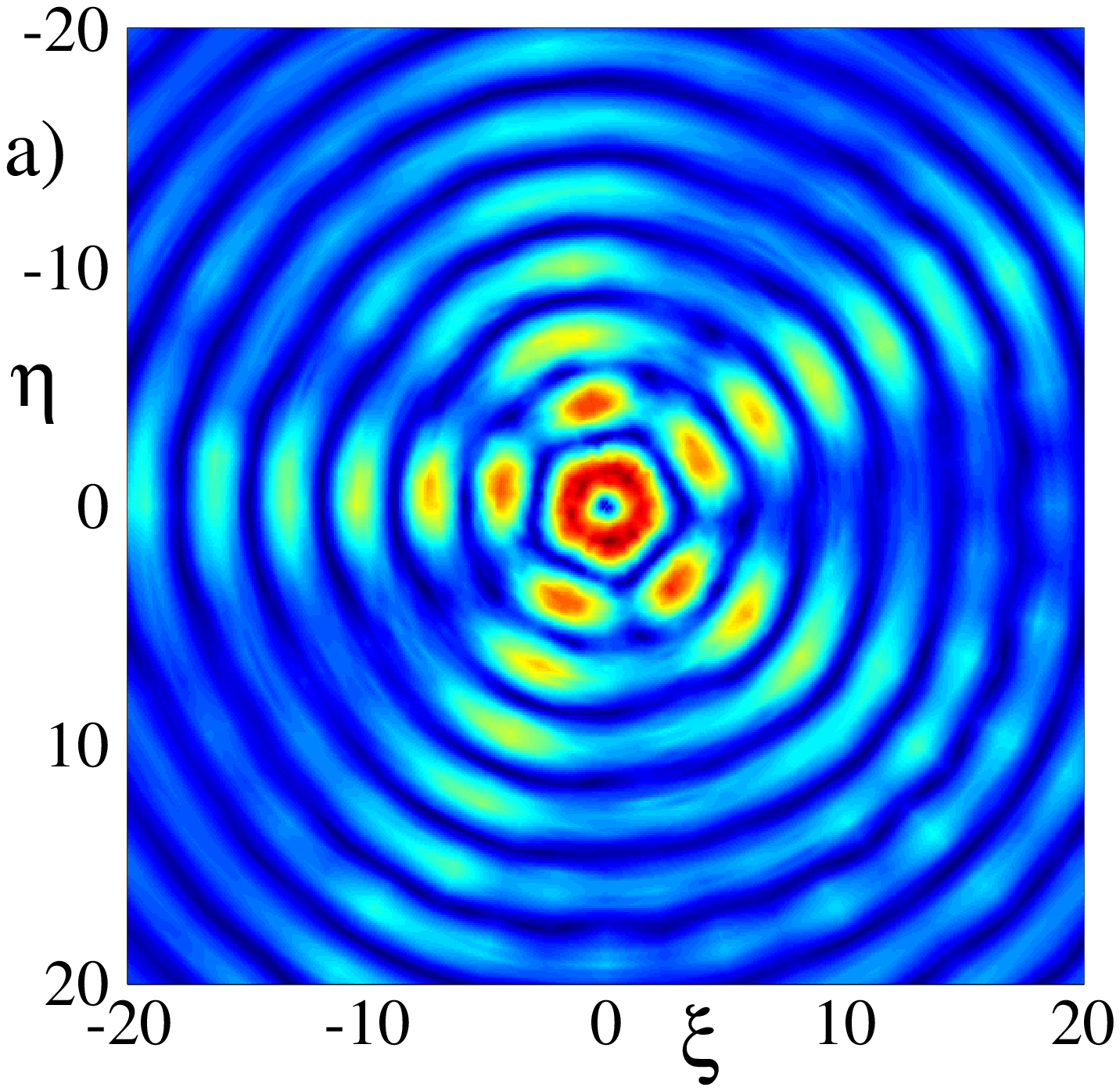} \includegraphics[width=4cm]{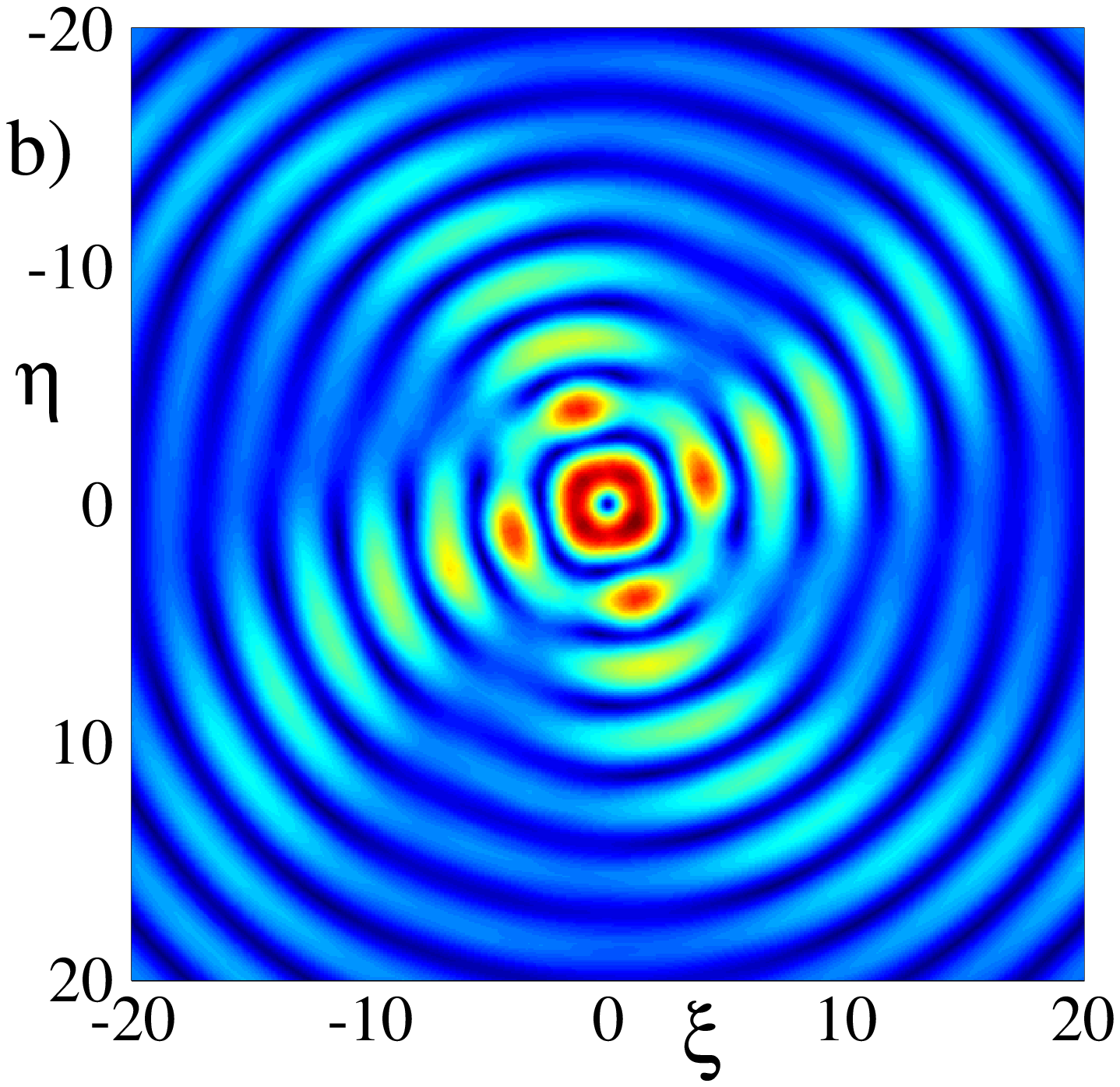}
\includegraphics[width=4cm]{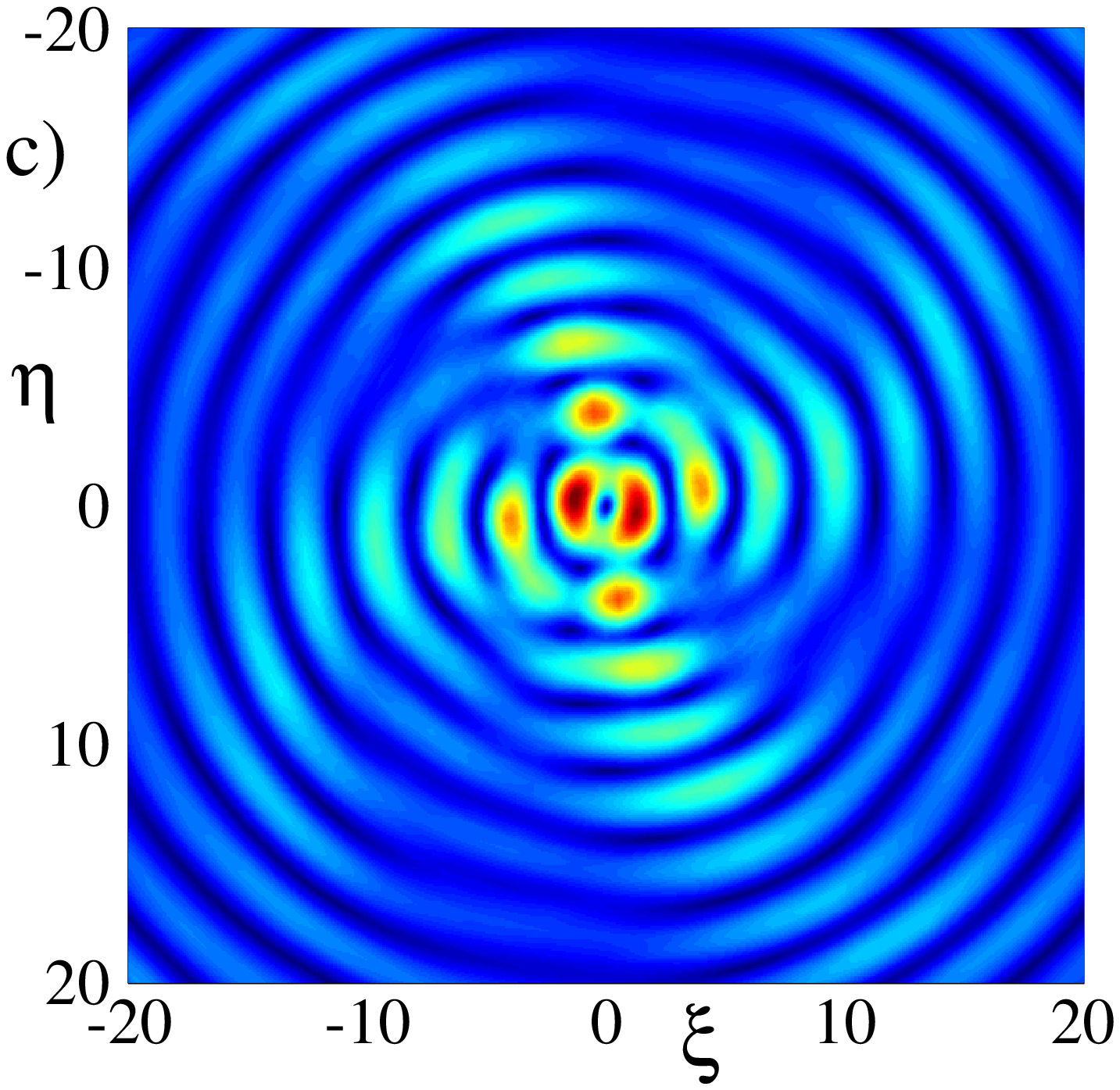} \includegraphics[width=4cm]{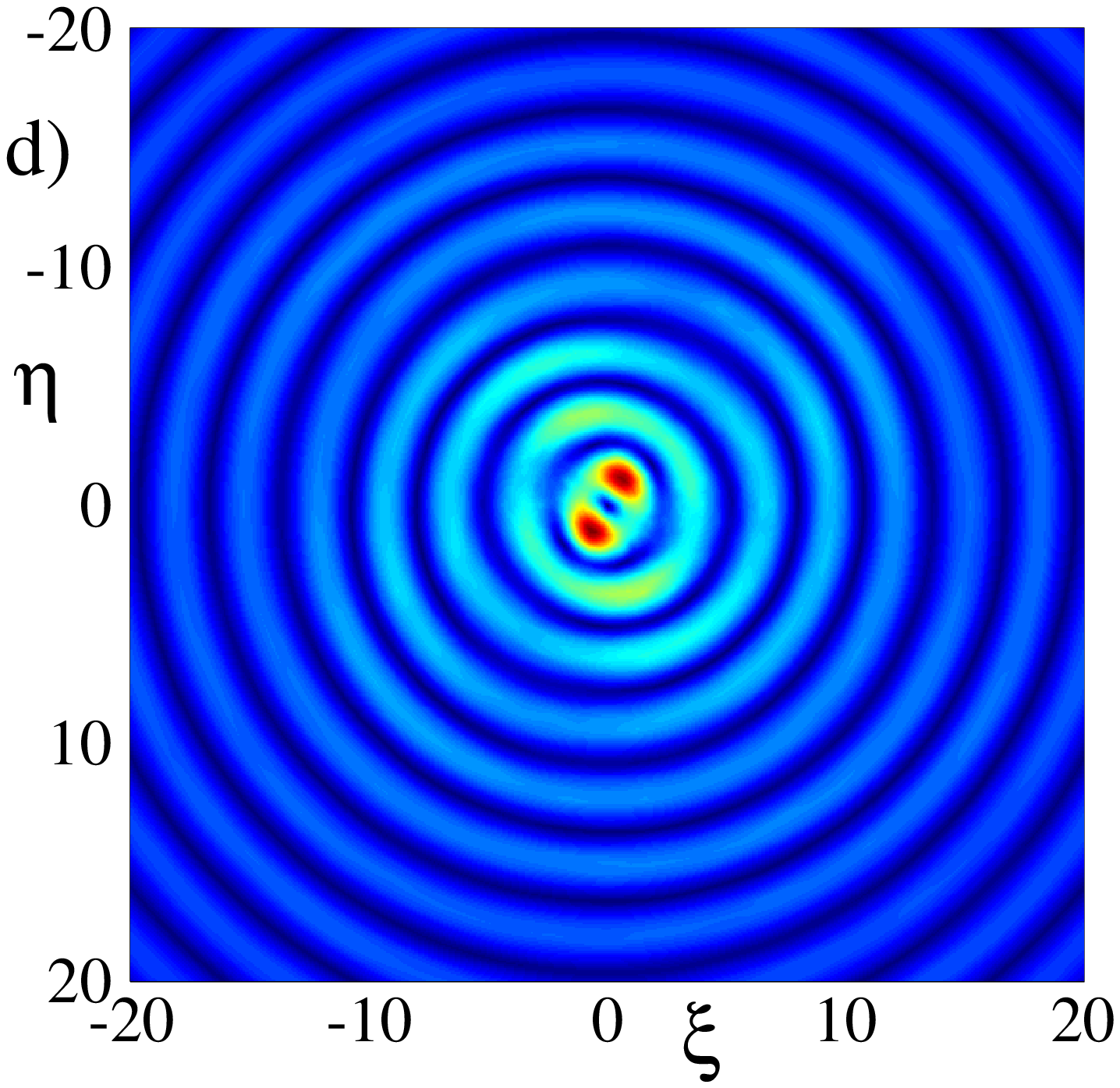}
\end{center}
\caption{For $M=4$, $s=1$ and $b_{0}=1.6$, the transverse intensity distributions, $|\tilde{A}|^{2}$, of the initially perturbed vortex with (a) $\protect\alpha =1.6$ at propagation distance $\protect\zeta =300$, (b) $\protect\alpha =2.2$ at $\protect\zeta =100$, (c) $\protect\alpha =2.8$ at $\protect\zeta =61.25$, and (d) $\protect\alpha =4$ at $\protect\zeta =15$. The scaled coordinates are $(\protect\xi ,\protect\eta )=\protect\sqrt{2k| \protect\delta|}(x,y)$. The numbers of fragments into which the unstable vortices split are exactly predicted by the linear-stability analysis.}
\label{fig3}
\end{figure}

We have further checked these predictions in direct simulations of Eq. (\ref{NLSE2}) with the nonlinear BVBs initially perturbed by random noise [using up to $3200\times 48$ points in the $(\rho,\varphi)$ plane]. In all the cases, rings of unstable BVBs tend to break into fragments moving along circular trajectories, whose number is exactly equal to the winding number $m$ of the largest instability growth rate, as can be seen by comparing Fig. \ref{fig3} to Fig. \ref{fig2}(b). The larger the instability growth rate, the sooner the BVB breaks. The mode competition observed in Fig. \ref{fig3}(c), with the main ring splitting in two fragments and the outer ones in four, implies that there are two perturbation eigenmodes which are nearly equally unstable.

\begin{figure}[tbp]
\begin{center}
\includegraphics[width=4cm]{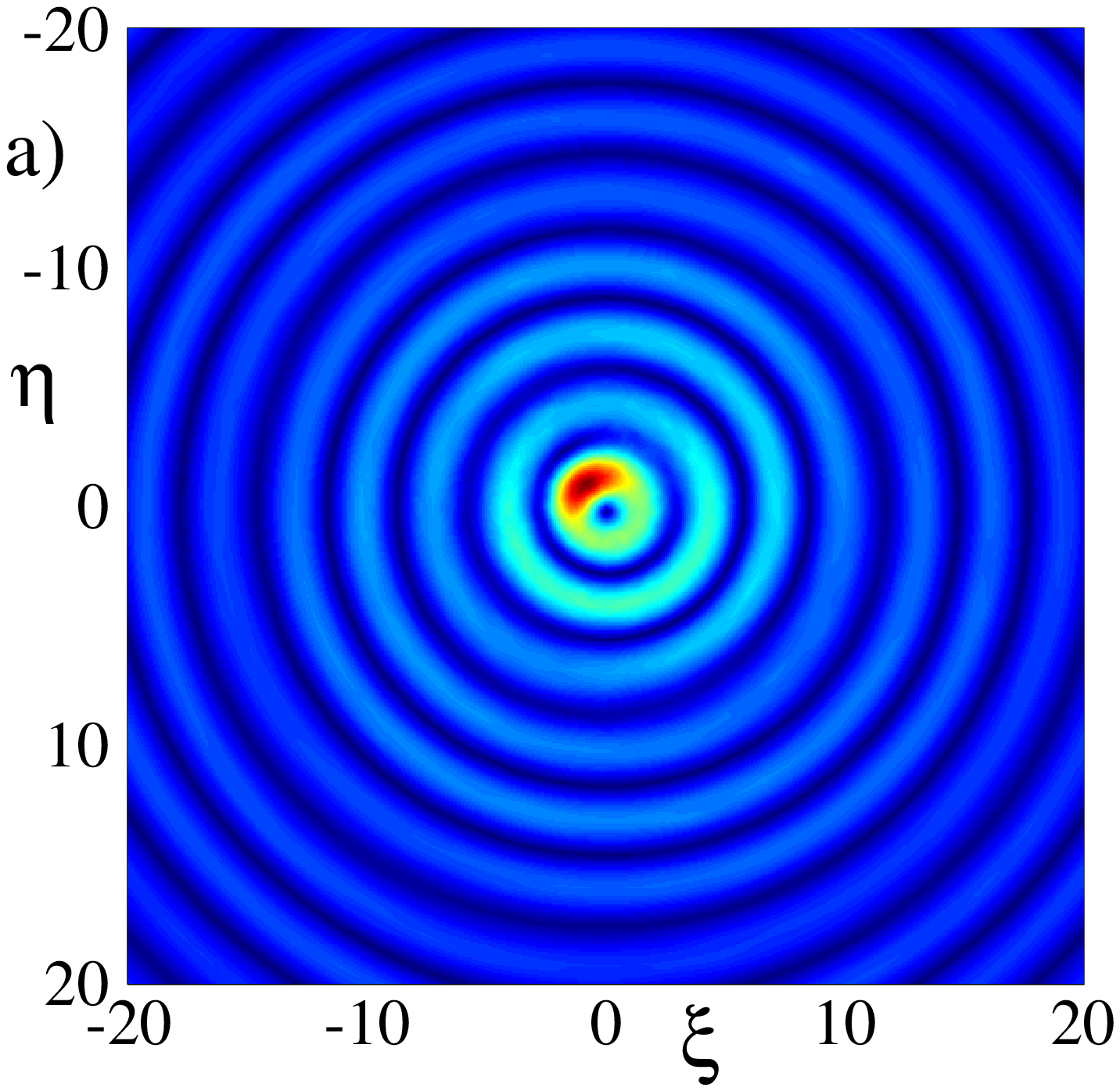} \includegraphics[width=4cm]{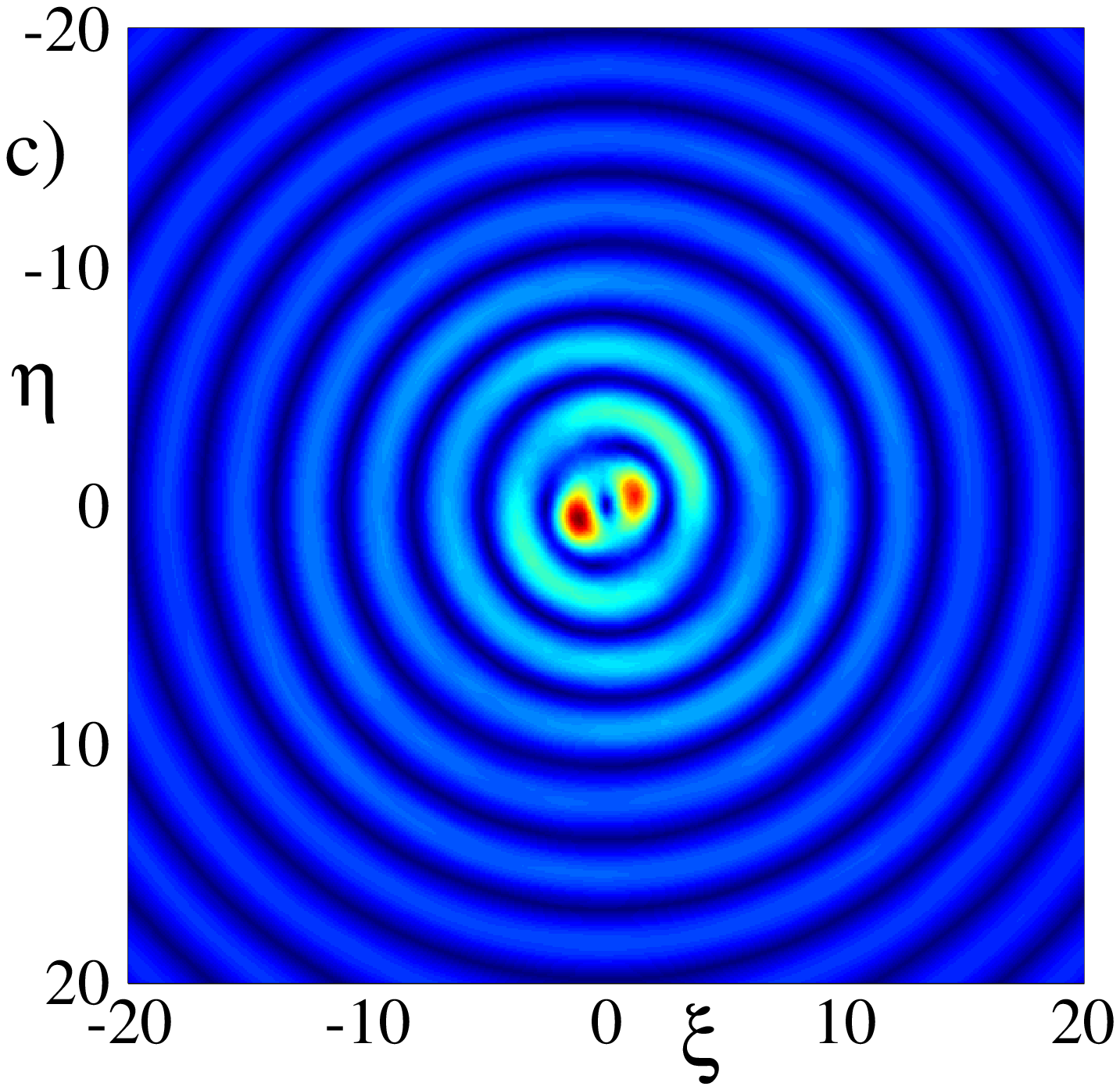}
\includegraphics[width=4cm]{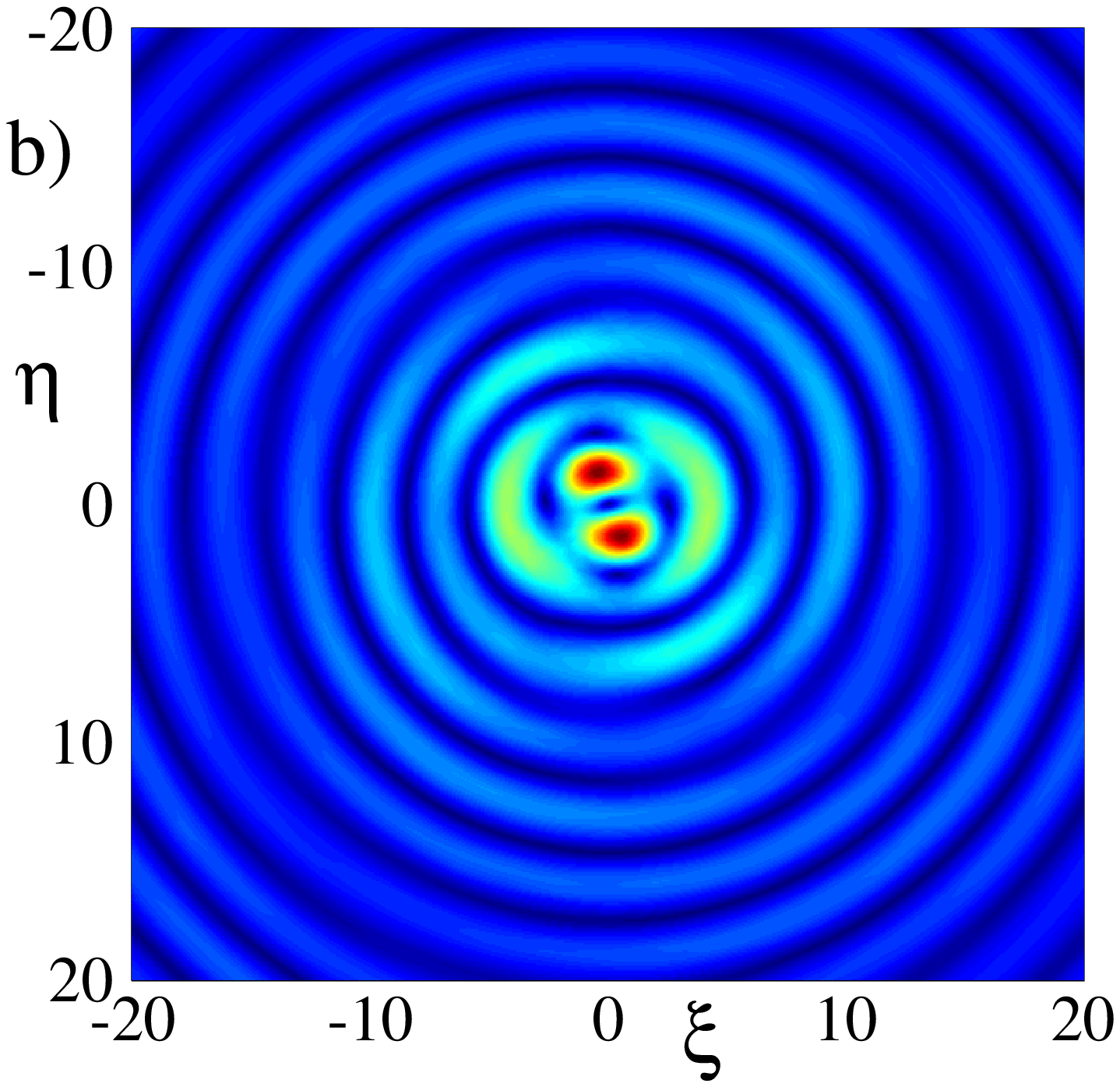} \includegraphics[width=4cm]{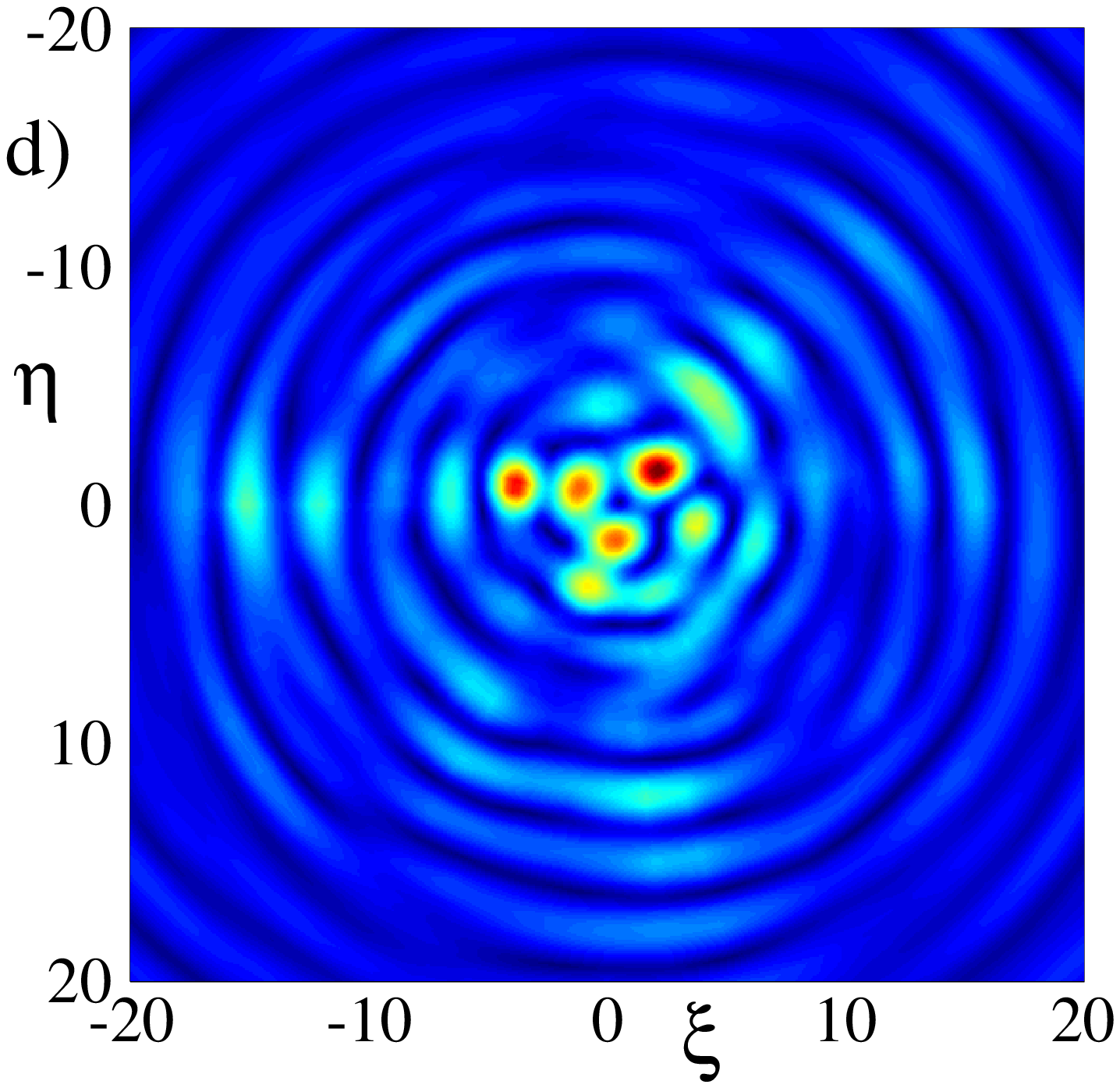}
\end{center}
\caption{For $M=4$, $s=1$ and $b_{0}=1.2$, the transverse intensity distributions, $|\tilde{A}|^{2}$, of the initially perturbed BVB with $\protect\alpha =3$ at propagation distances (a) $\protect\zeta =41.25$ and (b) $\protect\zeta =91.25$, and with $\protect\alpha =6$ at (c) $\protect\zeta =15$ and (d) $z=100$. These results demonstrate the secondary breakup of fragments produced by the primary instability of the vortices.}
\label{fig4}
\end{figure}

In contrast to the instability of the usual vortex solitons \cite{general-reviews}, the number of circulating fragments in the small-perturbation regime does not necessarily determine the outcome of the instability development for large perturbations. For example, the single [as predicted by Fig. \ref{fig2}(a)] circulating fragment in Fig. \ref{fig4}(a) breaks into two in Fig. \ref{fig4}(b), and the two [also as predicted by Fig. \ref{fig2}(a)] circulating fragments in Fig. \ref{fig4}(c) decay into numerous randomly placed splinters, which appear and disappear in the course
of the propagation, see Fig. \ref{fig4}(d). When the stability is predicted, the BVB absorbs the random perturbation and propagates, keeping its shape, as long as the simulations were run (up to $\zeta =400$).

\begin{figure}[t]
\begin{center}
\includegraphics[width=4cm]{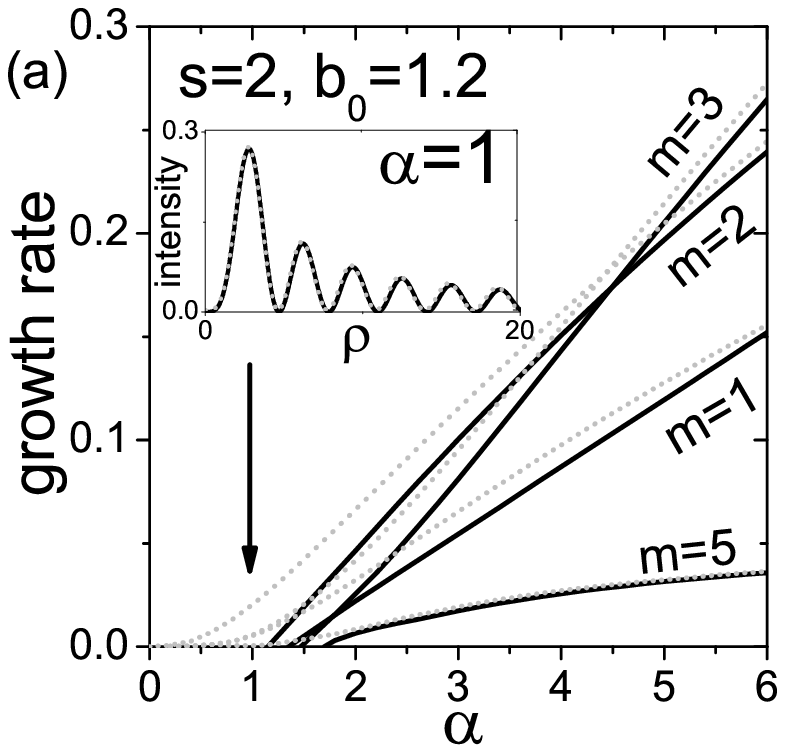} \includegraphics[width=4cm]{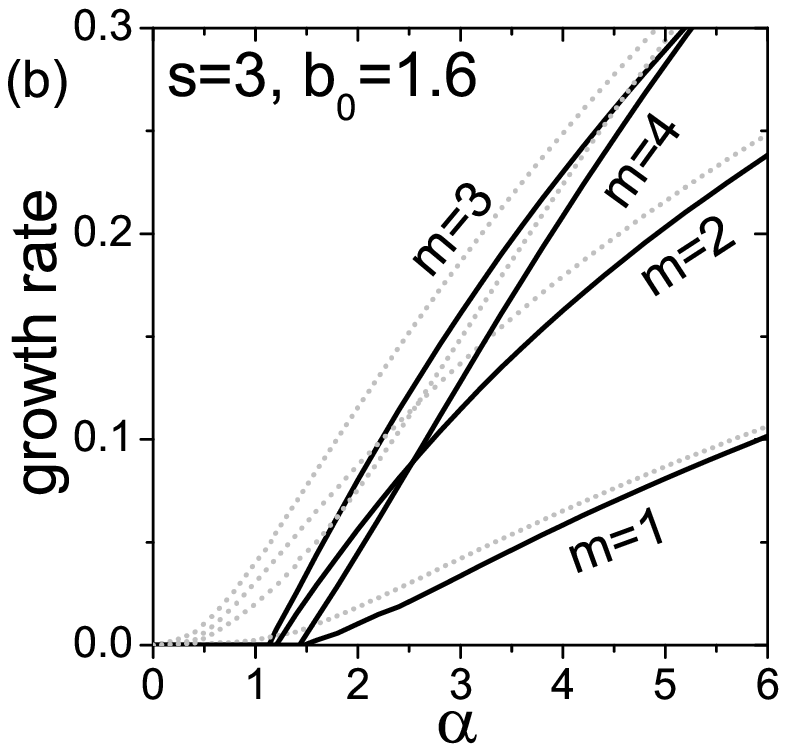}
\end{center}
\caption{For $M=4$, dimensionless growth rates of the most unstable azimuthal modes of BVBs (solid curves), and of their counterparts with the same parameters in the absence of absorption (dashed curves). Panels (a) and (b) display the results for $s=2$ and $3$, respectively. Inset in (a): dimensionless radial intensity profiles of the indicated stable (black curve) and unstable (gray dashed curve) BVBs in media with and without absorption, respectively.}
\label{fig5}
\end{figure}

Figure \ref{fig5} summarizes the results of the stability analysis for BVBs with $s=2$ and $s=3$, thus proving the existence of \textit{stable}
nonlinear BVBs with \textit{multiple vorticities}. A similar situation is expected to take place at $s>3$. Note that the previously elaborated
settings, such as those in Refs. \cite{2D,Dum,we,HP}, \textit{could not} stabilize vortices with $s>1$.

The mechanism that renders the vortices stable in the present system can be understood from the results in Fig. \ref{fig5}, where we have also plotted the instability growth rates for nonlinear BVBs in the fully transparent medium, obtained by setting artificially to zero all dissipative terms in
the equations. Naturally, the BVBs are always unstable in this case case. At $s=2$, $b_{0}=1.2$ and $\alpha =1$, for instance, the BVB intensity
profiles obtained in the models without and with MPA are almost identical [they cannot be distinguished on the scale of the inset in Fig. \ref{fig5}
(a)], although the former vortex is unstable, while the latter one is stabilized by the MPA.

In Refs. \cite{JUKNA} and \cite{XIE}, stabilization is reported to occur when the cone angle is increased, i.e., $\alpha $ is diminished, and this observation is qualitatively explained as suppressed growth of the modulation instability or nonlinear wave mixing at these cone angles. The
results of the linear-stability analysis in Fig. \ref{fig5} quantify the stabilizing effect of using large cone angles, both with and without MPA.
However, as seen in Fig. \ref{fig5}, the stabilization by increasing the cone angle is never complete if MPA is not considered, or, in other words, increasing the cone angle cannot, by itself, produce completely stable BVBs without MPA.

\section{Interpretation of the tubular, rotating and speckle-like filament regimes in axicon-generated nonlinear Bessel vortex beams}

Finally, we analyze the experimental implications of these results. We consider the usual arrangement in which a Gaussian beam with embedded
vorticity $s$ is passed through an axicon \cite{COUAIRON,PORRAS2,JUKNA}, or equivalent device to produce a conical beam \cite{XIE}. The optical field at the entrance of the nonlinear medium, placed immediately after the axicon, is assumed to be $A=\sqrt{I_G}\exp(-r^2/w_0^2)\exp(-i k\theta r)\exp(is\varphi)$ \cite{COUAIRON,SOCHACKI}, where $\sqrt{I_G}$ is the amplitude of the Gaussian beam, $w_0$ its width, and $\theta$ the cone angle imprinted by the axicon. In our dimensionless variables, this field reads as $\tilde{A}_{G}=b_{G}\exp (-\rho^{2}/\rho _{0}^{2})\exp (-i\rho )\exp (is\varphi )$, where amplitudes and lengths scale as in Eqs. (\ref{SCALING}). If the propagation after the axicon were linear, a finite-power version of the linear Bessel beam $\tilde{A}_{B}\simeq b_{B}J_{s}(\rho)\exp (is\varphi)$ of intensity $b_{B}^{2}=\pi \rho _{0}e^{-1/2}b_{G}^{2}$ \cite{COUAIRON,NOTE} would be formed at the axicon focus placed at $w_0/2\theta$ from the axicon, or in our dimensionless axial coordinate, at $\rho _{0}/4$, half of the Bessel distance $\rho _{0}/2$.

For the zero-vorticity configuration ($s=0$), it has been recently shown \cite{PORRAS2} that, in the course of the nonlinear propagation after the
axicon, the zero-vorticity nonlinear Bessel beam with the same cone angle as its linear counterpart, and with amplitude $b_0$ such that $|b_{\mathrm{in}}|=b_B$, determines the dynamics in the Bessel zone: If the nonlinear Bessel beam is stable, a steady regime is observed, characterized by smooth formation of the nonlinear beam and its decay afterwards; if it is unstable, an unsteady regime is observed in the Bessel zone, where the signatures of the development of the instability of the unstable nonlinear Bessel beam are clearly identifiable \cite{PORRAS2}.

Extensive numerical simulations allow us to conclude that similar phenomenology is valid for vortex beams ($s\neq 0$). The nonlinear BVB with
the same cone angle and vorticity as its linear counterpart, and with $b_0$ such that the amplitude of the inward H\"{a}nkel component is $|b_{\mathrm{in}}|=b_B$, determines the dynamics in the Bessel zone. This is demonstrated by the fact that the azimuthal-symmetry-breaking dynamics in the Bessel zone reproduces what is expected to follow from the stability or instability of that specific BVB. Three representative examples are shown in Fig. \ref{fig6}, and further details are reported in the supplement material \cite{SUP}. The three illuminating Gaussian beams, with embedded single vortex $s=1$, $\rho_0=400$, and $b_{G}=0.0402$ (left), $0.0341$ (center) and $0.030$ (right), would yield linear-Bessel amplitudes $b_{B}=1.11,0.94$ and $0.829$, respectively, in linear propagation. These values have been chosen such that, in media with $M=4$ and $\alpha=1,3$ and $6$, the nonlinear BVBs with $|b_{\mathrm{in}}|=b_B$ all three have $b_0=1.2$, as in Fig. \ref{fig2}(a) and Fig. \ref{fig4}. According to these figures, the BVB with $\alpha =1$ is stable, while ones with $\alpha =3$ and $6$ are unstable. In the two latter cases, the inverse growth rates, or characteristic length of development of the instability, are much shorter than the length of the Bessel zone, $\rho_0/2$. In the case of the stability (left panels in Fig. \ref{fig6} and more details in supplement \cite{SUP}), the BVB with $b_0=1.2$ and $\alpha=1$ is smoothly formed at the axicon focus, corresponding to distance $\rho_{0}/4=100$, which is followed by its smooth decay. In the unstable cases (central and right panels in Fig. \ref{fig6}, see additional details in the supplement \cite{SUP}), the dynamics within the Bessel distance reproduces that of the development of the instability leading to the azimuthal breakup, starting from the small-perturbation regime predicted by the linear-stability analysis in Fig. \ref{fig2}(a) and proceeding to the large-perturbation regime displayed in Fig. \ref{fig4}. At $\alpha =3$, a rotatory regime with one spot, further breaking into two, is observed, as also seen in Figs. \ref{fig4} (a) and (b). At $\alpha =6$, the instability is stronger, and the two rotating fragments quickly convert themselves into randomly placed non-rotating spots, as seen in Figs. \ref{fig4}(c) and (d).

These three situations closely resemble the quasi-stationary and rotary regimes described numerically in Ref. \cite{JUKNA}, and the
quasi-stationary, rotary and speckle-like regimes observed experimentally in Ref. \cite{XIE}. In those works, the regime which is associated with the
formation of a BVB is supposed to correspond to a stable BVB, and in Ref. \cite{JUKNA} the rotary regime is conjectured to be associated with either an unstable BVB or nonexistence of a BVB in the specific experimental configuration. In neither case the respective BVB was identified, therefore its stability cannot be studied to verify these conjectures. The above-mentioned representative examples, in the light of the linear-stability analysis, reveal that all the three regimes admit a common explanation in terms of the stability/instability of a specific nonlinear BVB, which is identified as one preserving the cone angle and topological charge, and amplitude $b_0$ such that $|b_{\mathrm{in}}|=b_B$. Since for the linear Bessel beam $b_{\mathrm{in}}=b_{\mathrm{out}}=b_B$, relation $|b_{\mathrm{in}}|=b_B$ states that the amplitude of the inward H\"ankel component is a third preserved quantity in the nonlinear dynamics. We also note from Ref. \cite{PORRAS3} that, for given $M$, $\alpha$ and $s$, $|b_{\mathrm{in}}|$ is a strictly growing function of $b_0$, from $|b_{\mathrm{in}}|=0$ at low $b_0$ up to $|b_{\mathrm{in}}|=\infty$ for $b_{0,\mathrm{max}}$ \cite{PORRAS3}. Thus, there always exists a single nonlinear BVB satisfying condition $|b_{\mathrm{in}}|=b_B$.

For example, in the experimental observation of the quasi-stationary, rotating and speckle-like regimes reported in Fig. 4 of Ref. \cite{XIE}, the
material constants, cone angle and three pulse energies allows us to identify the three BVBs as those defined by $M=5$, $s=3$, $\alpha=14.89$, and $b_0=0.368, 0.822$ and $1.644$. The stability analysis gives the respective largest dimensionless growth rates $0.026$, $0.245$ and $0.784$, or, multiplying by $|\delta|$, the growth rates $2.48$, $23.4$ and $74.9$ cm$^{-1}$ in physical units. Comparing the associated characteristic lengths of the instability development, $0.402$, $0.043$, $0.013$ cm, with the length $\sim 0.072$ cm of the Bessel zone explains the nonoccurrence of the azimuthal breakup as lack of development of the BVB instability in the first case, the rotating filaments in the second case as a primary manifestation of the breakup of the BVB, and the random filaments in the third case as full development of the instability.

Finally, the existence of truly stable nonlinear BVBs implies that a tubular-beam propagation regime for beams passing the axicon exists, being
limited solely by the finite amount of power stored in the reservoir, the depletion of which defines the edge of the Bessel zone. Thus, the stable
vortex tubules may be extended indefinitely long by increasing the power stock (e.g., by dint of the increase of $\rho_0$).

\begin{figure}[t]
\begin{center}
\includegraphics[width=8.5cm]{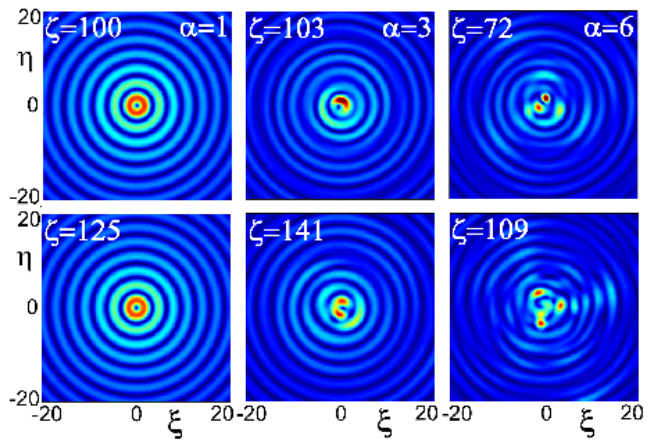}
\end{center}
\caption{Transverse intensity distributions, $|\tilde A|^2$, at the indicated propagation distances for $\protect\alpha =1$ (left), $\protect\alpha =3$ (center), and $\protect\alpha =6$ (right), in the presence of the four-photon absorption, produced by axicons illuminated by vortex-carrying Gaussian beams with $\protect\rho_{0}=400$ and $s=1$, but different amplitudes: $b_{G}=0.0402$ (left), $0.0341$ (center) and $0.030$ (right), yielding Bessel amplitudes $b_{B}=1.11,0.94$ and $0.829$, respectively. The three nonlinear BVBs with $s=1$ and $|b_{\mathrm{in}}|$ matching these values of $b_{B}$ in samples with the corresponding values of $\protect\alpha$ and $M$, have the same parameter $b_{0}=1.2$, which can be identified using the method of Ref. \protect\cite{PORRAS3}.}
\label{fig6}
\end{figure}

\section{Conclusions}

We have proposed a new setting for the stabilization of localized vortex beams in self-focusing Kerr media. The inclusion of the nonlinear
dissipative term accounting for the MPA (multiphoton absorption) makes the nonlinear Bessel vortex beams stable. As described earlier for the
zero-vorticity situation \cite{PORRAS1}, for Airy beams \cite{LOTTI}, and, recently, for nonlinear Bessel vortex beams \cite{PORRAS3,JUKNA}, the
concomitant MPA-induced power loss is balanced by the radial flux from their intrinsic reservoir. Stability regions for the nonlinear Bessel vortex beams have been predicted by the linear-stability analysis and corroborated by direct simulations. Stable states with multiple vorticities have been found too.

Differently from preceding works, these stable vortices do not require materials with ``tailored" nonlinearities. They may be common dielectrics,
such as air, water or standard optical glasses, all featuring the ubiquitous Kerr nonlinearity and MPA (multiphoton absorption) at sufficiently high
intensities (tens of TW/cm$^2$ in gases, or a fraction of TW/cm$^2$ in condensed matter). This is probably the reason why, on the contrary to
previous proposals, the practical implementation of BVBs (Bessel vortex beams) precedes the demonstration of their stability. Axicon-generated BVBs
with finite power were observed in \cite{XIE} to propagate steadily within the Bessel zone. The proof of the stability has important consequences for
these experiments and subsequent applications. The stability implies, for example, that the steady regime can be extended indefinitely long by simply
enlarging the Bessel zone.

The stability analysis and diagnostic numerical simulations also allow us to interpret the two other propagation regimes observed in these experiments. We have concluded that the nonlinear dynamics is always governed by a specific nonlinear BVB, which we have fully specified. The quasi-stationary regime in the Bessel zone is observed when this nonlinear BVB is found to be stable in our linear-stability analysis, or is effectively stable over the finite length of the Bessel zone. The rotary and speckle-like regimes are observed when this nonlinear BVB is found to be unstable, and the instability can develop, depending on its growth rate and the length of the Bessel zone. The azimuthal-symmetry-breaking dynamics in the Bessel zone is just a manifestation of the evolution of the unstable BVB.

As an extension of this analysis, it may be of interest to study the interaction between spatially separated stable vortices in the same model, as well as the stability of the tubular vortices against spatiotemporal (longitudinal) perturbations.

\acknowledgments

M.A.P. acknowledges support from Projects of the Spanish Ministerio de Econom\'{\i}a y Competitividad No. MTM2012-39101-C02-01, MTM2015-63914-P, and No. FIS2013-41709-P.


\begin{thebibliography}{99}
\bibitem{NOTE} This formula has been previously derived for the fundamental Bessel beam. Simple consideration based on the Fresnel diffraction and asymptotic methods show its validity for high-order Bessel beams.
\end{thebibliography}
\end{document}